\documentclass[letterpaper,useAMS]{mn2e}

\usepackage{graphicx,times}
\usepackage[totalwidth=480pt, totalheight=680pt]{geometry}

\def\deg{$^{\circ}$}

\def\kms{km s$^{\rm{-1}}$}

\title[Both velocity components recovered in barred discs]
{How to recover both velocity components in discs of barred 
galaxies with integral-field spectroscopy}

\author[Witold Maciejewski, Eric Emsellem and Davor Krajnovi\'{c}]
{Witold Maciejewski$^{1}$\thanks{E-mail:wxm@astro.livjm.ac.uk}, 
 Eric Emsellem$^{2,3}$ and Davor Krajnovi\'{c}$^{2}$\\
$^{1}$Astrophysics Research Institute, Liverpool John Moores University, Twelve
Quays House, Egerton Wharf, Birkenhead, CH41 1LD\\
$^{2}$European Southern Observatory, Karl-Schwarzschild-Str. 2, 85748
Garching, Germany\\
$^{3}$Universit\'e Lyon 1, Observatoire de Lyon, Centre de Recherche
Astrophysique de Lyon \\ \hspace*{0.5cm} and Ecole Normale Sup\'erieure de
Lyon, 9 avenue Charles Andr\'e, F-69230 Saint-Genis Laval, France}
\begin{document}

\maketitle

\begin{abstract}
We present a new method that derives both velocity components in the
equatorial plane of a barred stellar disc from the observed line-of-sight 
velocity, assuming geometry of a thin disc. The method 
can be applied to large departures from circular motion, and does not 
require multipole decomposition. It is based on assumptions that the 
bar is close to steady-state (i.e. does not evolve fast), and that
both morphology and kinematics are symmetrical with respect to the major 
axis of the bar. We derive the equations used in the method, and analyze
the effect of observational errors on the inferred velocity fields. 
We show that this method produces meaningful results via a simple toy model.
We also apply the method on integral-field data of NGC 936, for which
we recover both velocity components in the disc.
Knowing both velocity components in the disc, i.e. the non-observable 
transverse velocity in addition to the line-of-sight velocity, puts 
additional constraints on dynamical models and allows for new ways
of determining parameters that are crucial in characterizing galaxies.
\end{abstract}

\section{Introduction}
In the last decade, integral-field spectroscopy enabled unprecedented progress
in our understanding of the dynamics and evolution of galaxies (e.g. de Zeeuw 
et al. 2002, F\"{o}rster Schreiber et al. 2009). With spectra 
sampled throughout the entire galaxy, 
regular rotation of the galaxy can be verified and more complex motions
can be detected. The nature of these motions is usually recovered by
comparing the moments of the line-of-sight velocity distribution (LOSVD)
to dynamical models (e.g. Cappellari et al. 2006). However, the signatures
expected in the LOSVD may depend on the assumed model, and therefore 
model-independent signatures, or signatures that depend on minimal 
number of assumptions, are of great value. 

The dependence of the results on the assumed model can be seen on the 
example of the first moment of the LOSVD, i.e. the line-of-sight velocity,
analyzed for non-axisymmetric distortions in a planar flow in a galactic
disc. Harmonic analysis was applied to this problem by Franx et al. (1994)
and Schoenmakers et al. (1997), but it is limited to small departures from
circular motion, and interpreting amplitudes of various multipoles in terms
of different physical scenarios is unclear (Wong, Blitz \& Bosma 2004). Limiting the
expansion to only the initial multipoles distorts the results: when only 
the $m=1$ terms are included (radial and tangential velocity), often 
radial motions of large amplitude are implied (e.g. Swaters et al. 2003).
This leads to a continuity problem, as the radially traveling matter should
have a source or sink in the galaxy centre. Spekkens \& Sellwood (2007) 
showed that when an $m=2$ term is added (oval flow), it removes the need
for radial bulk motion. Moreover, adding this term changes considerably 
the derived rotation curve. The difficulty to distinguish radial bulk 
motion from oval flow has been noticed earlier for radio and long-slit data
(e.g. van der Kruit 1974, 1976; Bosma et al. 1977).
Thus one needs to characterize non-axisymmetric distortions in the 
observed kinematics in a way that can be more directly related to 
the dynamical state of the galaxy.

For a galactic disc, the line-of-sight (LOS) velocity constitutes only one 
out of the two velocity components. The other component is the transverse 
velocity in the disc, which generally cannot be determined with the current 
observing techniques for discs of external galaxies.
However, if a galaxy possesses some symmetry, or a steady
pattern is present there, one can use these properties to derive the 
transverse component. This has been done by Sridhar \& Sambhus (2003), but
it required the knowledge of the angular speed of the rotating pattern,
which often cannot be determined with high certainty.

Here we present a new method to derive both velocity components in a disc 
galaxy that hosts a bar or a dominant bisymmetric perturbation in the 
stellar component. The two velocity components are recovered from the LOS
velocity using the symmetry introduced by the bar, and since the galactic 
disc needs to be deprojected in this method, the disc inclination and 
the position angle of its line of nodes must be known. Unlike the previous 
attempt to model non-circular motions in oval distortions to the gravitational
potential (Spekkens \& Sellwood 2007), our method is not limited to
multiplicity $m \le 2\;$ --- in fact it returns the complete
radial and tangential velocity components throughout the disc without the
need for multipole expansion. From the recovered streaming motion one can 
constrain the bar's contribution to the overall gravitational potential.
By applying 
the continuity equation directly to the two velocity components, one can 
estimate the pattern speed of the bar.

In Sect.2, we derive the equations through which we recover, from the
orientation of the galaxy and the LOS velocity, the radial and tangential
velocity components. In Sect.3, we analyze how the recovered values are
affected by errors in the adopted orientation of the galaxy. We show
how the method presented here works in practice by deriving in Sect.4 the
radial and tangential velocity fields in the disc of NGC 936, from the
SAURON integral-field data (Krajnovi\'{c} et al. 2011; Emsellem et al. 2012). 
In Sect.5, we discuss 
how knowing both components of the velocity field puts better constraints
on the structure and dynamics of disc galaxies.

\section{The method}
In this method, we assume that: (1) the bar is in steady-state (it does not 
evolve fast), (2) the bar is symmetrical with respect to its major
axis, (3) the disc of the galaxy is flat, so that there are only two 
velocity components.
The first assumption is well justified by our understanding of
evolution of barred galaxies (e.g. Shen \& Sellwood 2004). The
second one possibly breaks at the ends of the bar, where often a
parallelogram-like morphology is seen, particularly 
when a transition to spiral arms is present. The third assumption may
not be valid if the centre of the galaxy is dominated by a bulge.
Thus the assumptions of this method are best fulfilled in the middle
of the radial extent  of the bar. The perturbation caused by the bar is 
also strongest there.

Imagine a star moving in the disc plane on the trailing side of the bar 
with velocity $v_{B1}$ in the frame rotating with the bar (Fig.1, where we 
assume that the bar is rotating counter-clockwise). This motion is 
described by Newtonian dynamics, hence reversing time gives equally 
possible motion. Upon time reversal, the bar starts rotating clockwise, 
the star is now on its leading side, and its velocity arrow changes 
direction (marked by the thick dashed line in Fig.1). If the bar is 
symmetrical with respect to its major axis, this state upon time reversal 
is identical to the state on the other side of the bar's major axis before 
time reversal, i.e. to the star moving with velocity $v_{B2}$. Thus in a 
symmetric bar, stars moving on one side of the major axis of the bar with 
velocities $v_{B1}$ should be mirrored by stars moving on the other side 
of the bar with velocities $v_{B2}$. This is true regardless of what orbit 
the star is on: regular symmetric or not symmetric, or irregular.
These orbits are likely to be similarly populated on both sides of the bar, 
as otherwise the appearance of the bar would not be symmetrical with respect 
to its major axis. Therefore the {\it mean} velocity on one side of the
bar should be mirrored by that on its other side, and this is the basis 
of our method. However, in order to emphasize that the star can move from 
one side of the bar to the other, we connect the two positions in Fig.1 by 
a simple oval orbit, keeping in mind that our analysis is not limited to 
such orbits.

Because of the symmetry analyzed above, Cartesian components of the 
velocity vectors in the frame rotating with the bar, in the plane of
the disc, at two locations symmetrical with respect to the bar's 
major axis are related by
\begin{equation}
v_{xB2} = -v_{xB1} \;\;\;\;\;\;\;\;\;\;\;\;\; v_{yB2} = v_{yB1},
\end{equation}
where the $B$ index indicates that the $x_B$ axis is aligned with the 
major axis of the bar. When converted to polar 
coordinates, the components are related by
\begin{equation}
v_{R2} = -v_{R1} \;\;\;\;\;\;\;\;\;\;\;\;\; v_{\varphi 2} = v_{\varphi 1},
\end{equation}
with the coordinate system now not dependent on the position of the bar.

If we want to relate the two velocity components in the disc of a galaxy, 
that is projected on the sky, to the observed LOS velocity, we may
impose the Cartesian coordinates in the inertial frame of the disc
in such a way that the line-of-nodes (LON) coincides with the $x$-axis.
Then only the $y$-component of the velocity in the disc will contribute 
to the observed LOS velocity, and the two are related by\footnote{in 
contrast to velocities 
measured in the frame rotating with the bar, velocities in the galaxy disc 
measured in the inertial frame are marked with prime}
\begin{equation}
v_{LOS} = v'_y \sin i,
\end{equation}
where the galaxy disc is inclined to the plane of the sky at angle $i$.
The $x$-component will remain unobservable, constituting exclusively 
the transverse velocity (i.e. the proper motion), whose vector remains in 
the plane of the sky.

\begin{figure}
\centering
\rotatebox{90}{\includegraphics[width=0.5\linewidth]{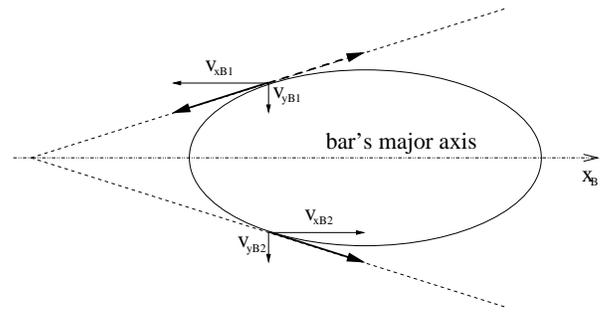}}
\caption[]{For a star moving in the plane of a barred galactic disc, 
Cartesian velocity components, in the frame rotating with the bar, 
are shown at two locations, symmetrical to the 
bar's major axis (the $x_B$ axis). The thick dashed arrow indicates
the velocity vector upon time reversal (see Sect.2).}
\label{f1}
\end{figure}

In the method proposed here, we use the deprojected velocity field $v'_y$,
which can be obtained for the known inclination of the galaxy $i$ and the 
position angle (PA) of its LON. Then for a disc that contains a steady-state 
asymmetry, like a bar, we can recover both
velocity components using the relation (2) between two independent 
components at two locations symmetrical with respect to the major axis
of the bar. It is possible, because at each of the two locations, the
two velocity components contribute differently to $v'_y$. In Fig.2, the
geometry used in deriving formulae for the two velocity components is
clarified. The bar is located at an arbitrary angle $\alpha$ to the LON 
in the plane of the galaxy disc\footnote{$\alpha$ is the angle of 
counterclockwise rotation from the positive half of the LON to the bar's 
major axis}.
In the frame rotating with the bar, contributions from the polar velocity 
components $v_R$ and $v_{\varphi}$ to the Cartesian velocity component $v_y$ 
are
\begin{equation}
v_{y1} = v_{\varphi 1} \cos(\alpha - |\gamma|) + v_{R1} \sin(\alpha - |\gamma|),
\end{equation}
at position 1, and 
\begin{equation}
v_{y2} = v_{\varphi 2} \cos(\alpha + |\gamma|) + v_{R2} \sin(\alpha + |\gamma|),
\end{equation}
at position 2. The angle $\gamma$ is measured in the plane of the galactic 
disc, and spans between the bar's major axis 
and the line connecting each of the two symmetrically located positions, 
at which $v'_y$ is inferred from the LOS velocity. 

The $v_y$ velocity 
components in (4) and (5) are in the frame rotating with the bar. They 
can be converted to the inertial frame by adding to $v_{\varphi}$ 
the term $\Omega_B R$, where $\Omega_B$ is the angular velocity of the bar, 
and $R$ is indicated in Fig.2. Thus the $y$-velocity components
in the inertial frame, which directly relate to the LOS velocity through (3),
can be written as
\begin{eqnarray}
v'_{y1} & = & (v_{\varphi 1}+ \Omega_B R) \cos(\alpha - |\gamma|) + v_{R1} \sin(\alpha - |\gamma|)\\
v'_{y2} & = & (v_{\varphi 2}+ \Omega_B R) \cos(\alpha + |\gamma|) + v_{R2} \sin(\alpha + |\gamma|)
\end{eqnarray}
As $v_{R2}$ and $v_{\varphi 2}$ are related to $v_{R1}$ and $v_{\varphi 1}$ by (2),
substituting (2) to (7) gives
\begin{equation}
v'_{y2} = (v_{\varphi 1}+ \Omega_B R) \cos(\alpha + |\gamma|) - v_{R1} \sin(\alpha + |\gamma|).
\end{equation}
Now there are only two unknowns in the system of two equations (6) and (8):
the radial and tangential velocities $v_{R1}$ and $v_{\varphi 1}$, for which
the system can be readily solved, yielding
\begin{eqnarray}
v_{R1} & = & \frac{v'_{y1} \cos(\alpha + |\gamma|) - v'_{y2} \cos(\alpha - |\gamma|)}{\sin(2\alpha)}\\
v_{\varphi 1} & = & \frac{v'_{y1} \sin(\alpha + |\gamma|) + v'_{y2} \sin(\alpha - |\gamma|)}{\sin(2\alpha)} - \Omega_B R
\end{eqnarray}
Thus if the galaxy is symmetrical with respect to the major axis of its bar,
one can recover both velocity components in the disc plane {\it in the 
inertial frame}: $v_{R1}$ and $v_{\varphi 1} + \Omega_B R$. Note that for this
we use the {\it deprojected} velocity field $v'_y$, hence the galaxy has to
be deprojected before applying this method. In Sect.4, we show how this 
procedure works in practice. Because of the
singularity in the denominator in (9) and (10), this method fails when 
$\sin(2\alpha)=0$, i.e. for the bar either parallel or perpendicular to the
LON. One can expect best results for the bar at moderate angles to the LON.

\begin{figure}
\centering
\includegraphics[width=\linewidth]{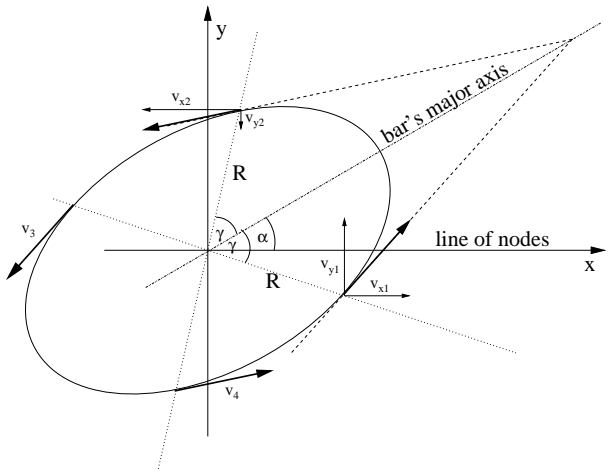}
\caption[]{Same as Fig.1, but the LON is now along the Cartesian 
$x$-coordinate, and the bar's major axis is at angle $\alpha$ in the galactic 
plane to the LON. With this choice of coordinates, only the
$y$ velocity component contributes to the observed LOS velocity,
while the $x$ component is unobservable, remaining in the plane of the sky.
This diagram is drawn in the plane of the galactic disc.}
\label{f2}
\end{figure}

We visualize the expected results by showing in Fig.3 the radial and 
tangential velocity components for a cartoon model of an elliptical flow. 
In arbitrary units, this model assumes a rotation curve that rises up to 
the radius of 200, and is flat further out at the amplitude of 200 
arbitrary units. The motion is circular outside the radius 
of 400, but it becomes elliptical inside that radius, which is intended to
mimic the motion in a bar. The elliptical motion reaches maximum 
eccentricity at semi-major axis $a=200$, where the axis ratio is $b/a=0.5$. 
The motion further inside is becoming
increasingly circular, which intends to simulate the area dominated by the
bulge. The elliptical flow intended to follow the bar rotates like a solid
body with respect to the inertial frame, with corotation radius at 400.
Three example oval trajectories, including the most eccentric one, are 
overplotted in Fig.3.

\begin{figure}
\centering
\includegraphics[width=\linewidth]{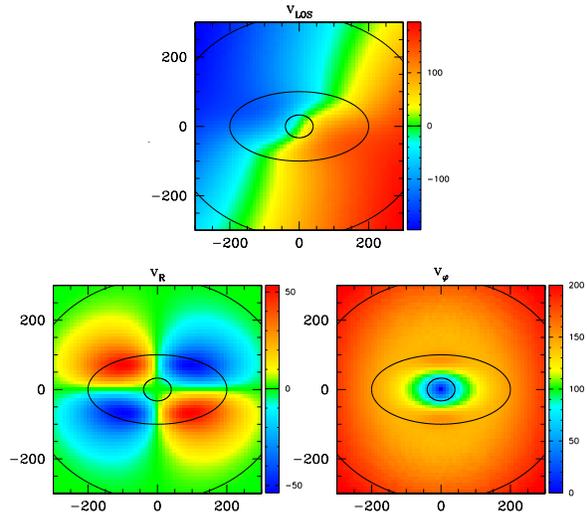}
\vspace{-45mm}
\caption[]{For a cartoon model of an elliptical flow following a bar, 
defined at the end of in Section 2, we show 
the deprojected LOS velocity $v'_y$ ({\it top}), 
as well as the radial ({\it bottom-left}) and tangential ({\it bottom-right}) 
velocities in the disc plane. Three example oval trajectories
are overplotted. The colour scale is shown in the attached wedges.
For PA reference, north is up.}
\label{f3}
\end{figure}

The top panel of Fig.3 displays the deprojected LOS velocity $v'_y$,
when the PA\footnote{PA is measured east of north, to the positive half of 
the LON in kinematics} of the bar is 90\deg, and that of the LON is 
240\deg. One can notice an asymmetry caused by the elliptical flow. Its most 
evident signature can be seen in the radial velocity component (bottom-left
panel of Fig.3) in 
the form of four velocity extrema in four quadrants: two maxima and two
minima, alternately. The signature is strong: in our cartoon model with
the maximum eccentricity of the flow being $b/a=0.5$, it has an amplitude
of 54 for the rotation curve at the amplitude of 200 in its flat part. 
As expected, the signal is strongest where the flow is most
eccentric. The tangential velocity component, shown in the bottom-right 
panel of Fig.3, displays complex structures that have no straightforward 
correspondence to the structure of the flow. By experimenting with
various additional cartoon models, we noticed that these structures 
strongly depend on the variation of the eccentricity of the flow and on the 
adopted rotation curve, while the structure of the radial velocity component,
shown in the bottom-left panel of Fig.3, remains virtually unchanged.
As expected, using equations (9) and (10) we can recover the radial and 
tangential velocity fields in the lower panels of Fig.3  from the 
deprojected LOS velocity $v'_y$ in the top panel of Fig.3.

\section{Effects of observational errors}
In addition to the accuracy in measurement of the LOS velocity itself, 
the structure and the amplitude of the derived two velocity components
in the disc plane can be
affected by observational errors in the systemic velocity of the galaxy,
the inclination of the disc, as well as in the position angles of the
bar's major axis and of the LON.

\begin{figure}
\centering
\includegraphics[width=\linewidth]{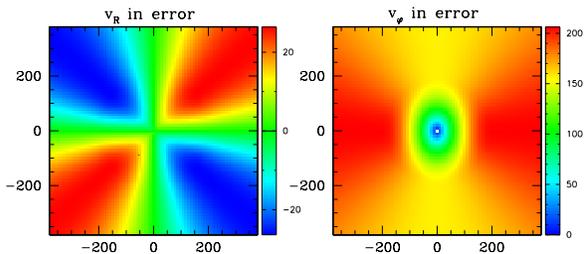}
\vspace{-85mm}
\caption[]{Erroneous signatures in the derived radial ({\it left}) and 
tangential ({\it right}) velocity fields in the disc plane, caused by 
adopting the PA of the LON being 5\deg\ 
in error, while the actual motion is purely circular with the rotation 
curve same as in Fig.3. The velocity fields were derived using (9) and (10)
with the alleged bar at $\alpha=20$\deg\ to the adopted LON, and horizontal
in these plots. The colour scale is shown in the attached wedges.}
\label{f4}
\end{figure}

\subsection{Error in the PA of the LON}
The potentially most problematic source of error is the uncertainty on the PA
of the LON, as one signature that it creates is similar to that of the oval
flow. In Fig.4, we show erroneous radial and tangential velocity fields 
that our method produces for a disc in purely circular rotation, when
the PA of the LON is 5\deg\ in error. We adopted here the same rotation
curve as in Sect.2. If the PA of the LON were correct, the
tangential velocity field should be axisymmetric,
while the radial velocity field should be uniformly zero. 

The erroneous signature in the radial velocity field in the left panel of 
Fig.4 resembles the signature of real oval flow in the bottom-left panel of 
Fig.3. Both are of bisymmetric nature, with opposite extrema in each
quadrant. However, there is one major difference between these signatures:
while the signature of a real oval flow is strongest at the radius where the 
flow is most eccentric (Fig.3, bottom-left panel), signature of the misplaced 
LON continues to arbitrarily large radii (Fig.4, left panel). Therefore
in order to assure that the radial motion recovered with our method is
genuine, it is essential that in addition to sampling the region where 
the flow is expected to be non-circular, the region outside is sampled 
as well.

The magnitude of the error in the derived radial velocity changes with 
the PA of the bar relative to the LON (the $\alpha$ angle in our method), 
being largest when the bar is parallel or perpendicular to the LON. The 
error in the left panel of Fig.4 is showed for $\alpha=$20\deg, for which
it reaches the magnitude of 26 (for reference, the flat part of the rotation 
curve in the same units is at 200). The error has the same pattern and 
magnitude when the bar is 20\deg\ from the direction perpendicular to the
LON, i.e. for $\alpha=$70\deg. For $\alpha=$10\deg\ or 80\deg, the magnitude
reaches 51, while for $\alpha=$40\deg\ or 50\deg, it is reduced to 15.
Clearly, the radial velocity field is least affected by the error in the 
PA of the LON when the angle $\alpha$ between the LON and
the bar is around 45\deg\ or 135\deg.

When the tangential velocity field is derived with our method for purely 
circular rotation, but with the PA of the LON in error, it is no longer
axially symmetric, but it develops 
another bisymmetric pattern, now with two maxima and two minima along 
the bar's major axis and perpendicular to it (Fig.4, right panel).
Like in the case of the derived radial velocity, the magnitude of the 
error in the derived tangential velocity changes with $\alpha$, and is
minimized for $\alpha$ around 45\deg\ or 135\deg. In Fig.4, the derived 
tangential velocity field is shown for $\alpha=$20\deg. In the flat part
of the rotation curve, where it should be constant, it varies between 156
and 206. Thus the amplitude of variation is 25, similar to that in the
radial motion, and the average tangential velocity is 181, smaller than 
the rotation curve in the flat part, which is 200. When $\alpha=$10\deg,
the amplitude of variation in the derived tangential velocity increases 
to 47, and its average is at 157. The same variation is for 
$\alpha=$80\deg, but then the average is 240. Thus a bisymmetric pattern
in the derived tangential velocity field similar to that in the right
panel of Fig.4 can serve as indicator of an incorrect PA of the LON adopted
for this derivation. As can be seen in Fig.3, it should not accompany a 
genuine streaming observed in radial motion.

\subsection{Inclination and bar PA errors}
The error in the disc inclination propagates from (3) to (9) and (10) as a 
multiplicative constant, hence it affects the amplitude of both radial
and tangential motion, as expected. Incorrect deprojection of the disc 
can also result in spurious signatures in radial and tangential velocity 
field similar to those in Fig.4, even if the disc is in purely circular 
motion and the LON is identified correctly. However, the amplitudes
of these signatures are considerably smaller than those of the signatures
indicating an error in the position angle of the LON that we described 
in Sect.3.1.
The error in inclination must be at least 30\deg\ in order to generate 
the amplitude of signatures the same as in the case of the PA of the 
LON being 5\deg\ in error.

In order to estimate the error in the derived velocity field caused
by adopting an incorrect PA of the bar, we derived the radial and 
tangential velocity fields for the oval motion described in Sect.2, but
for the PA of the bar in the disc plane being 5\deg\ in error. The 
structure of the derived radial velocity field, shown in the bottom-left 
panel of Fig.3, remains virtually unaffected by this error:
its bisymmetric structure remains the same, and its amplitude is 
altered only by 5\%. The structure also gets stretched along
the bar, but its aspect ratio does not change by more than 
10\%. Most importantly, contrary to the error in the PA of the LON, which
produces signatures in the radial velocity field that extend to arbitrarily 
large radii (Fig.4, left panel), such signatures are absent when the PA of 
the bar is in error.

The structure of the tangential velocity field
changes moderately because of the PA of the bar being in error: the rate of 
growth along the bar, and the magnitude of the local maximum
on bar's minor axis change by a factor less than 10\%. Like the radial 
velocity field, the tangential velocity field is affected by an error in the 
bar's PA only within the bar. At larger radii it is axisymmetric, even when
the PA of the bar is in error. No
bisymmetric structure like that from the right panel of Fig.4 is seen,
which indicates that such a structure is a unique signature of an incorrect
PA of the LON.

\subsection{Error in the systemic velocity of the galaxy}
The error  $\epsilon$ in the systemic velocity, when $v'_{y1} + \epsilon$ and 
$v'_{y2} + \epsilon$ are being used in (9) and (10) instead of $v'_{y1}$ and 
$v'_{y2}$, propagates to the derived radial and tangential velocity 
components in such a way that the erroneous values, $v^\epsilon_{R1}$ and 
$v^\epsilon_{\varphi 1}$, relate to the correct ones by
\begin{eqnarray}
v^\epsilon_{R1}       & = & v_{R1} \; - \; \epsilon \frac{\sin |\gamma|}{\cos \alpha}\\
v^\epsilon_{\varphi 1} & = & v_{\varphi 1} \; + \; \epsilon \frac{\cos |\gamma|}{\cos \alpha}.
\end{eqnarray}
As $v_{R2}=-v_{R1}$, this generates an $m=1$ mode in the radial velocity field.
However, this effect can be completely removed by invoking the symmetry
with respect to the {\it minor} axis of the bar. In Fig.2, we used two
LOS velocities, $v'_{y1}$ and $v'_{y2}$, at two locations, $\alpha \pm \gamma$,
symmetric with
respect to the major axis of the bar, in order to obtain the two velocity 
components in the disc, $v_{R1}$ and $v_{\varphi 1}$. Similarly, at locations 
$\alpha + \pi \pm \gamma$, i.e. on the other side of the minor axis of the 
bar, one can derive the two velocity components, $v_{R3}$ and $v_{\varphi 3}$
from the LOS velocities $v'_{y3}$ and $v'_{y4}$. If the flow is symmetric
with respect to the minor axis of the bar then these components should be 
equal to $v_{R1}$ and $v_{\varphi 1}$. The $v_{R3}$ and $v_{\varphi 3}$ velocity 
components can be derived from equations analogous to (9) and (10):
\begin{equation}
v_{R3} \; = \; \frac{v'_{y3} \cos(\alpha + \pi + |\gamma|) - v'_{y4} \cos(\alpha + \pi - |\gamma|)}{\sin(2(\alpha + \pi))}
\end{equation}
\begin{equation}
v_{\varphi 3} \; = \; \frac{v'_{y4} \sin(\alpha +\pi + |\gamma|) + v'_{y4} \sin(\alpha + \pi - |\gamma|)}{\sin(2(\alpha +\pi))} - \Omega_B R.
\end{equation}
If $v'_{y3}$ and $v'_{y4}$ are affected by the same error in the systemic 
velocity $\epsilon$ as $v'_{y1}$ and $v'_{y2}$ are affected above, the error
propagates to $v_{R3}$ and $v_{\varphi 3}$ yielding erroneous values
$v^\epsilon_{R3}$ and $v^\epsilon_{\varphi 3}$ that relate to the correct ones by
\begin{eqnarray}
v^\epsilon_{R3}       & = & v_{R3} \; - \; \epsilon \frac{\sin |\gamma|}{\cos (\alpha+\pi)} \; = \; v_{R3} \; + \; \epsilon \frac{\sin |\gamma|}{\cos \alpha}\\
v^\epsilon_{\varphi 3} & = & v_{\varphi 3} \; + \; \epsilon \frac{\cos |\gamma|}{\cos (\alpha+\pi)} \; = \; v_{\varphi 3} \; - \; \epsilon \frac{\cos |\gamma|}{\cos \alpha}.
\end{eqnarray}
By comparing (15) and (16) to (11) and (12) one can see that the error in 
$v_{R3}$ and $v_{\varphi 3}$ is of the same magnitude, but of opposite sign 
compared to that in $v_{R1}$ and $v_{\varphi 1}$. Since from symmetry 
$v_{R3}=v_{R1}$ then averaging $v^\epsilon_{R1}$ and 
$v^\epsilon_{R3}$ cancels the error originating from the incorrect systemic 
velocity.  The same is true for $v_{\varphi 3}$ and $v_{\varphi 1}$.
In both cases, the symmetrization can be written as
\begin{equation}
v^{\rm symmetrized} (x,y) = \frac{v(x,y) + v(-x,-y)}{2} .
\end{equation}

Certainly the price for this averaging is the loss of information about 
possible difference in velocity structure on the two sides of the minor 
axis of the bar. However, if this difference is not caused by inaccurate 
derivation, but is rather intrinsic to the galaxy itself, then it is 
likely that the velocities are not symmetrical with respect to the {\it major}
axis of the bar, either. Thus the method presented here is rather unlikely
to recover differences in kinematical structure at the two ends of the bar.
Note that the symmetrization is the last, additional step of our method, after 
the radial and tangential velocities have been determined independently on 
the two sides of the minor axis of the bar, and therefore it does not
affect the method. It can be performed optionally, if one wants to remove the 
effects of error in systemic velocity.

\subsection{Summary of effects of observational errors}
In summary, the error in the PA of the LON is the strongest source of
uncertainty in the derived velocity fields.
However, this error can be easily spotted, as it generates a unique 
signature of bisymmetry in the derived tangential velocity (Fig.4, right
panel), and its signature in the derived radial velocity field (Fig.4, left
panel) is different at large radii from the signature of genuine streaming 
motion (Fig.3, bottom-left panel). 
Errors in the inclination and in the PA of the
bar affect the derived velocity field in lesser degree, while the error
in the systemic velocity has no effect if the velocity field is symmetrized
with respect to the minor axis of the bar.

\section{Application to the SAURON IFU data of NGC 936}
NGC 936 is a barred S0 galaxy, whose morphology and kinematics have been 
studied in detail in the pre-IFU era (Kormendy 1983, 1984; Kent 1987; Kent
\& Glaudell 1989, Merrifield \& Kuijken 1995). It is very luminous and hence
massive, which helps maximizing the measured LOS velocities. Because of
the strong bar present in NGC 936, the amplitude of non-circular motions
is expected to be large as well. The bar in NGC 936 is not nearly 
aligned with either principal axis, which minimizes errors caused by
adopting inaccurate geometry, as analyzed in Sect.3. 

\begin{table}
\caption{Parameters of NGC 936}
\begin{tabular}{lllll}
\hline
                             & K83       & KG89    & A3D       & adopted\\
\hline
recession velocity [\kms]    & --        & --      & 1429      & 1429\\
PA of photometric major axis & 136\deg   & 135\deg & 130.7\deg & --\\ 
PA of kinematic major axis   & --        & --      & 318.0\deg & 318\deg\\
inclination (0\deg\ for face-on) & 40.9\deg  & 41\deg  & 39\deg& 39\deg\\
PA of the bar on the sky     & 77\deg    & 79\deg  & --        & 81\deg\\
$\alpha$ between LON and bar & 114.4\deg & --      & --        & 117\deg\\
\hline
\end{tabular}

\medskip
Abbreviations for the sources of parameter values are: K83 -- Kormendy (1983,
note: inclination converted from a convention 0\deg\ for {\it edge-on} there),
KG89 -- Kent \& Glaudell (1989), A3D -- Cappellari et al. (2011), Krajnovi\'{c}
et al. (2011)
\end{table}

Here, we apply the method described in Sect.2 in order to recover 
the radial and tangential velocities from the integral-field data of 
NGC 936 observed with the SAURON integral-field spectrograph
(Bacon et al. 2001) mounted on the William Herschel telescope in La Palma.
In total, 40 exposures were obtained, covering the full extent of the bar, 
during a run of 5 consecutive nights in August 2003, with a few additional
exposures in January 2004. The data were
reduced as in Emsellem et al. (2004), and all exposures merged into a single
mosaic datacube. These data have been obtained in the course of a programme
to observe barred galaxies with SAURON (PI. Emsellem), and the resulting
mosaic has been used in the course of the ATLAS$^{\rm 3D}$ project 
(Cappellari et al. 2011) as NGC936 is part of that sample. The maps 
presented here (see Krajnovi\'{c} et al. 2011) will also be part of a 
paper presenting the mosaic in greater detail (Emsellem et al. 2012). 
Note that the bright point-like source present on the
South-Eastern side of the centre of the galaxy is a Supernova (2003gs)
discovered by Robert Evans which was still visible during the first run in
August 2003. Parameters of NGC 936 derived in previous studies and adopted 
here are listed in Table 1. Below we list the steps in which the observed 
two-dimensional LOS velocity array was processed in order to derive the 
two velocity components in the disc of the galaxy.

\begin{figure}
\centering
\includegraphics[width=\linewidth]{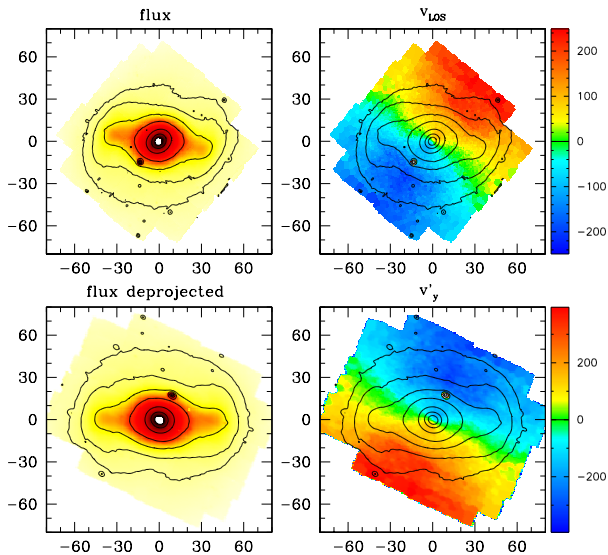}
\vspace{-46mm}
\caption[]{{\it Top-left:} The image of NGC 936 taken from the zeroth moment
of the full mosaic SAURON datacube. {\it Top-right:} The LOS velocity
(the first moment) of NGC 936 from the full mosaic SAURON datacube, after 
subtraction of the systemic velocity. North is up in the upper panels.
{\it Bottom-left:}
The image of NGC 936, deprojected using parameters from Table 1, and
rotated so that the bar is horizontal. {\it Bottom-right:} The LOS velocity,
deprojected and rotated in the same way as the image, and scaled by $\sin 
i$,
so that it results in the $v_y'$ velocity component in the disc plane of 
NGC 936, perpendicular to the LON. The bright point-like source is the
Supernova 2003gs. Isophotes are overplotted in contour.
The colour scale is shown in the attached 
wedges. The linear scale is in arcsec in the upper panels, and in 
arcsec deprojected to the disc plane in the lower panels.}
\label{f5}
\end{figure}

Firstly, the systemic velocity was subtracted from the observed LOS velocity.
The systemic velocity was derived from the heliocentric recession velocity
of 1429 \kms listed in the ATLAS$^{\rm 3D}$ survey (Cappellari et al. 2011),
after subtracting the heliocentric correction of 24 \kms. Thus the 
systemic velocity in the data is 1405 \kms.

Secondly, in order to obtain $v_y'$ in the galaxy plane from (3), the LOS
velocity field was deprojected, assuming the PA of the LON at 138\deg\ 
(318\deg\ for its positive half) and the disc inclination of 39\deg.
All the sources listed in Table 1 are in good agreement about the
inclination value. On the other hand,
the ATLAS$^{\rm 3D}$ estimate of the PA of the photometric major axis 
(Krajnovi\'{c} et al. 2011) gives a value 4\deg-5\deg\ smaller than earlier
estimates (Kormendy 1983; Kent \& Glaudell 1989), while
ATLAS$^{\rm 3D}$ estimate of the PA of the kinematic major axis is 
2\deg-3\deg\ larger than those estimates. We adopt here the kinematic 
estimate, acknowledging that it may be in error by up to 5\deg. 
We explored consequences of this uncertainty with similar magnitude 
in Sect.3.1. 

Thirdly, the derived $v_y'$ velocities at locations symmetrical 
with respect to the bar's major axis were substituted to (9) and (10) in 
order to recover $v_{R}$ and $v_{\varphi}$. In order to use this symmetry,
we tried to rotate the bar to the horizontal position using the PA of the bar
given by Kormendy (1983) and Kent \& Glaudell (1989), but this was resulting
in a slightly non-horizontal bar. We had to modify the PA of the bar to
81\deg\ in order to alleviate it, and this is the PA that we 
consequently use. Our adopted angle between the positive half of the LON 
and the bar on the sky 
is therefore $\alpha_{\rm sky}=$ 123\deg. After deprojection, the angle 
between the LON and the bar {\it in the plane of the galaxy disc} is 
$\alpha=$ 117\deg, as $\tan \alpha = \tan \alpha_{\rm sky} / \cos i$.

\begin{figure}
\centering
\includegraphics[width=\linewidth]{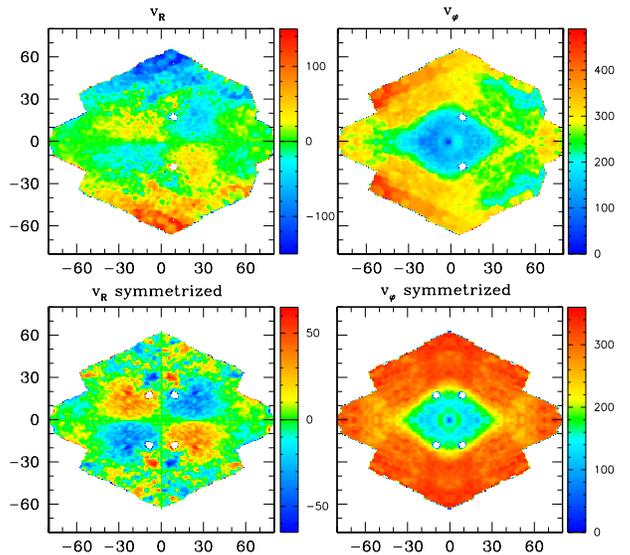}
\vspace{-46mm}
\caption[]{{\it Top-left:} The radial velocity field in the disc of NGC 936,
derived from (9). {\it Top-right:} The tangential velocity field derived 
from (10). {\it Bottom-left:} The radial velocity field symmetrized around 
the minor axis of the bar (after averaging velocities on the left and right 
of the minor axis of the bar). {\it Bottom-right:} The tangential velocity 
field, symmetrized in the same way. The point-like sources are mirror images
of the supernova. The bar is horizontal. The colour scale is shown in the 
attached wedges. The linear scale is in arcsec deprojected to the 
disc plane.}
\label{f6}
\end{figure}

The integrated luminosity derived from the full mosaic datacube is shown
in the top-left panel of Fig.5. In the top-right panel, we show the LOS
velocity field (the first moment of the SAURON data), after correction 
for the systemic velocity. The deprojected image of NGC 936, rotated so 
that the bar is horizontal, is shown in the bottom-left panel of Fig.5. 
The same deprojection and rotation was applied to the LOS velocity field.
In addition, the deprojected LOS velocity was divided by $\sin i$ in order 
to obtain the velocity component perpendicular to the LON in the disc plane 
of NGC 936 ($v_y'$ velocity in our method). This $v_y'$ velocity field, 
shown in the bottom-right panel of Fig.5, is used as the input in the
method presented here.

The derived radial and tangential velocity fields are shown in the upper 
panels of Fig.6. The fields are smaller than the deprojected velocity 
field from the bottom-right panel of Fig.5, because they could be created 
only when information from both sides of the bar is available.
In the region dominated by the bar, the pattern in the 
derived radial velocity of NGC 936 resembles the cartoon model from the 
bottom-left panel of Fig.3: it is bisymmetric ($m=2$), with maxima and 
minima in the quadrants defined by the bar. The appearance of the extrema 
is somewhat different on the two sides of the minor axis of the bar, with
their shapes stretched along the bar on the left, and more confined on the
right. This may hint at somewhat different kinematics at the two ends of the
bar in NGC 936. Outside the bar, there
is a clear $m=1$ pattern with a strong minimum at the top and a maximum at 
the bottom. In Sect.3.3, we showed that an $m=1$ pattern can be caused by an
error in the systemic velocity. Kinemetry fit (Krajnovi\'{c} et al. 2006) to 
the LOS velocity field of NGC 936 shows that the mean velocity rises almost 
monotonically in the inner 60 arcsec by 25 \kms. Therefore adopting 
for the systemic velocity the mean velocity in the inner parts of the galaxy, 
as we did here, results in a velocity asymmetry for the outer parts. Our 
method generates a typical signature of $m=1$ multiplicity in the outer parts, 
seen in the top-left panel of Fig.6. Averaging over the two ends of the bar 
naturally removes this effect, and it is performed below. The tangential 
velocity field (Fig.6, top-right) well recovers the rise of the rotation curve
to about 250 \kms\ in the inner 30 arcsec, but it is highly asymmetric
at larger radii. Upon closer inspection, antisymmetry with respect to the
minor axis of the bar dominates there, hence averaging over the two ends
of the bar should return a more reliable velocity field, and it is 
performed below.

The symmetrized radial and tangential velocity fields, after symmetrization 
using (17), are shown in the lower panels of Fig.6.
The symmetrization of the radial velocity field removes strong extrema
of $m=1$ multiplicity that were present at the top and bottom of this
field, and now the bisymmetric pattern dominates the field, having 
largest amplitude within the bar. The pattern does not continue to 
larger radii, hence it is a genuine signature of streaming motion, and not
of an error in the PA of the LON. The amplitude of radial velocity reaches
60 \kms\ locally, but after averaging over small fluctuations, it is in 
the range 30-40 \kms. When compared to the amplitude of the flat part of
the rotation curve of about 300 \kms\ (see below), this relative amplitude 
is 2-3 times smaller than in the cartoon model from Sect.2. Outside the 
bar, there is a hint of another, weaker bisymmetric pattern of opposite
sign at large radii. It may be related to change in the direction 
of major axis of orbits outside the corotation of the bar, in the outer 
disc, or it may indicate that the PA of the LON or the inclination is in 
slight error. In any case, this signature is much weaker (below 10 \kms) 
than the dominating signature of streaming motion in the bar.

The azimuthal dependence of the derived tangential velocity, seen in 
the top-right panel of Fig.6, has been largely removed after the field
has been symmetrized around the minor axis of the bar (Fig.6, bottom-right
panel). Now the symmetrized tangential velocity field agrees well with 
the cartoon model: it not only recovers the rise of the rotation
curve in the inner 30 arcsec, but also its flat part at larger radii,
where it reaches amplitude of about 300 \kms. However, the symmetrized
tangential velocity still shows some azimuthal variation at large radii
between 320 \kms\ near the minor axis of the bar and 280 \kms\ near the
major axis, with one feature as low as 260 \kms. This is consistent with
the PA of the LON or the inclination being in slight error, but that does 
not dominate the recovered signal.

\section{Discussion}
Transverse velocity can be observed as proper motion only for stars in our 
Galaxy and nearby objects. This motion is too small to be observed in external 
galaxies besides our nearest neighbours. 
However, if some degree of symmetry is assumed in the system, 
this unobservable velocity component can be deduced from the LOS velocity
and the galaxy orientation in space. The method presented here involves 
the symmetry of the flow with respect to the major axis of the bar.
It assumes geometry of flat disc, hence it neglects the vertical dimension
in the disc. Inclusion of a vertical spatial extent, like in a thick disc or 
a bulge, will introduce projection effects, important for highly inclined
discs. Inclusion of a vertical velocity component will not change the results
presented here as long as there is no vertical bulk motion, so that random
motions contributing to the vertical velocity dispersion dominate. Vertical 
bulk motion can affect the radial and tangential velocity components derived
with the method presented here, but it can exist only when the disc warps from
a plane. Warps within bars have not been observed, but similar effect can
be caused by a buckling bar, which has been theoretically predicted 
(Binney 1981; Combes \& Sanders 1981).
In Sect.2, we noted that this method assumes a steady-state bar, which excludes
a quickly evolving buckling bar. In a separate paper devoted to the
analysis of the integral-field spectra of NGC 936 (Emsellem et al. 2012),
we will build N-body models of this galaxy, and will search for significant
vertical bulk motions.

The method presented here can be applicable to stars, because of time 
reversibility in dissipationless Newtonian dynamics, but not
to gas, which is dissipational. Gas flows on the two sides of
the major axis of the bar differ considerably (e.g. Athanassoula 1992),
and with our method one cannot retrieve two velocity components 
of gas flow in a bar, and thus explicitly derive the inflow rate. 
Below we present implications of knowing the
two-dimensional {\it stellar} velocity field.

\subsection{Possible advantages of knowing both velocity components}
Knowing both velocity components in the disc clearly provides better
constraints on dynamical models. The method presented here recovers 
these components in a non-parametric way, without multipole expansion.
It has been shown (Spekkens \& Selwood 2007) that limiting the number
of terms in the multipole expansion can greatly alter the
rotation curve, in addition to predictions of significant radial flow 
that causes the 'continuity problem'.

The only explicit way to derive angular velocity of a rotating pattern,
the Tremaine-Weinberg method (Tremaine \& Weinberg 1984), relies on the 
continuity equation 
applied to the fluid of stars moving in a flat disc of a galaxy, in which
a pattern rotating with a well defined angular velocity is present. The
continuity equation in the plane of the disc involves, at any location, 
surface density of the tracer, as well as both its velocity components.
As only one velocity component can be observed, Tremaine and Weinberg
performed apt integrations, so that the dependence on the unobserved velocity
component can be removed, and in effect, the angular velocity of the pattern 
derived. Because of the need for such integration, the Tremaine-Weinberg
method involves integrals extending to infinity, but it has been shown that
it gives reliable estimates when integrals include the majority of the 
galactic emission (e.g. Aguerri, Debattista \& Corsini 2003).

Once we recover {\it both} velocity components in a disc of a barred 
galaxy with our method, the continuity equation can 
be solved for pattern speed at any location independently, without the need
of integrating this equation. This should allow studies of spatial 
dependence of pattern speed similar to those of radial dependence
(Merrifield, Rand \& Meidt, 2006; Meidt et al. 2008). However, as can be seen 
in Fig.6, the derived velocity field is noisy, and it must be spatially
averaged before any conclusions on pattern speed can be made. We are 
currently searching for the most efficient scheme of averaging.

Gravitational potential of the bar exerts torques that induce
non-circular motion of stars. Previously, bar torque was estimated 
e.g. from near-IR photometry (Laurikainen \& Salo 2002; Speltincx, 
Laurikainen \& Salo 2008) or by comparing the simulated velocity 
fields of the gas component to the observed H$_\alpha$, CO or HI velocity 
fields (Kranz, Slyz \& Rix 2003; Boone et al. 2007). This requires a 
large suite of hydrodynamical models and/or
estimates on mass-to-light ratio. In our method, constraints on the 
bar contribution to the total 
gravitational potential can be derived from the amplitude of the 
streaming motion in the bar, in particular the amplitude of the radial 
motion. This amplitude, normalized to the rotation curve,
is proportional to the bar torque, normalized by the total radial force.
The radial force characterizes the total gravitational potential, while
torque is likely to come exclusively from the bar. Thus in this way we
can estimate the bar contribution that is independent of the mass-to-light 
ratio, which is the source of bias in the photometric estimates.

\section{Conclusions}
We have presented a new method to derive both velocity components in a 
barred stellar disc from the observed LOS velocity. The method relies on 
symmetry of the flow around the major axis of the bar, but it is readily 
generalized for flows having other azimuthal periodicities (like lop-sided 
distributions). The method does not use multipole expansion and is not 
limited to small departures from circular motion. It provides best results
when the bar is not close to being parallel or perpendicular to the LON, and 
the integral-field data cover the whole bar, extending somewhat beyond it.
The values of the two derived velocity components are most affected by an
error in the PA of the LON, but if the derived values are in error, a clear 
signature appears in the derived velocity field. We applied the method to
the SAURON integral-field data of NGC 936 (Krajnovi\'{c} et al. 2011; 
Emsellem et al. 2012) and recovered patterns in the radial and tangential 
velocity field, which we expected based on a cartoon model of a simple 
generic oval flow. The method can be readily applied to 
barred galaxies observed in recent surveys, which cover the whole extent 
of the galaxy disc, like the VENGA (Blanc et al. 2010) and CALIFA 
(Sanchez et al. 2012) surveys. Future
wide-field integral-field units, like MUSE (Laurent et al. 2010), 
will provide data ideally
suited for this method. Having both velocity components in the disc 
puts better constraints on dynamical models, and opens new ways to
determine the mass and the pattern speed of the bar.

An Interactive Data Language implementation of the method presented 
in this paper is
available at the web address: http://www.eso.org/$\sim$dkrajnov/idl.
WM wishes to thank the European Southern Observatory for supporting this 
work during his research stay as a Visiting Fellow in November-December 
2011 at ESO Headquarters in Garching/Germany.

\section*{REFERENCES}
Aguerri, J., Debattista, V.~P.~\& Corsini, E.~M.\ 2003, MNRAS, 338, 465\\
Athanassoula, E.\ 1992, MNRAS, 259, 345\\
Bacon, R.\ et al.\ 2001, MNRAS, 326, 23\\
Binney, J.\ 1981, MNRAS, 196, 455\\
Blanc, G.~A.\ et al.\ 2010, ASP Conerence Series, Vol. 432, p.180\\
Boone, F.\ et al.\ 2007, A\&A, 471, 113\\
Bosma, A., van der Hulst, J.~M., \& Sullivan, W.~T.\ 1977, A\&A, 57, 373\\
Cappellari, M.\ et al.\ 2006, MNRAS, 366, 1126\\
Cappellari, M.\ et al.\ 2011, MNRAS, 413, 813\\
Combes, F. \& Sanders, R.~H.\ 1981, A\&A, 96, 164\\
de Zeeuw, P.~T.\ et al.\ 2002, MNRAS, 329, 513\\
Emsellem, E.\ et al.\ 2004, MNRAS, 352, 721\\
Emsellem, E.\ et al.\ 2012, in preparation\\
F\"{o}rster Schreiber, N.\ et al.\ 2009, ApJ, 706, 1364\\
Franx, M., van Gorkom, J.H. \& de Zeeuw, T.\ 1994, ApJ, 436, 642\\
Kent, S.~M.\ 1987, AJ, 93, 1062\\
Kent, S.~M.~\& Glaudell, G.\ 1989, AJ, 98, 1588\\
Kormendy, J.\ 1983, ApJ, 275, 529\\
Kormendy, J.\ 1984, ApJ, 286, 132\\
Krajnovic, D.\ et al.\ 2006, MNRAS, 366, 787\\
Krajnovic, D.\ et al.\ 2011, MNRAS, 414, 2923\\
Kranz, T., Slyz, A., \& Rix, H.-W.\ 2003, ApJ, 586, 143\\
Laurent, F.\ et al.\ 2010, SPIE Conference Series, 7739, 147\\
Laurikainen, E. \& Salo, H.\ 2002, MNRAS, 337, 1118\\
Meidt, S.E.\ et al.\ 2008, ApJ, 676, 899\\
Merrifield, M.~R.~\& Kuijken, K.\ 1995, MNRAS, 274, 933\\
Merrifield, M.~R., Rand, R.J. ~\& Meidt, S.E.\ 2006, MNRAS, 366, L17\\
Sanchez, S.F.\ et al.\ 2012, A\&A, 538, A8\\
Schoenmakers, R.H.M., Franx, M. \& de Zeeuw, P.T.\ 1997, MNRAS, 292, 349\\
Shen, J. \& Sellwood, J. A.\ 2004, ApJ, 604, 614\\
Spekkens, K. \& Sellwood, J. A.\ 2007, ApJ, 664, 204\\
Speltincx, T., Laurikainen, E. \& Salo, H.\ 2008, MNRAS, 383, 317\\
Swaters, R.A.\ et al.\ 2003, ApJ, 587, L19\\
Sridhar, S.~\& Sambhus, N.\ 2003, MNRAS, 345, 539\\
Tremaine, S.~\& Weinberg, M.~D.\ 1984, ApJ, 282, L5\\
van der Kruit, P.~C.\ 1974, ApJ, 188, 3\\
van der Kruit, P.~C.\ 1976, A\&A, 52, 85\\
Wong, T., Blitz, L., \& Bosma, A.\ 2004, ApJ, 605, 183

\end{document}